\documentclass[preprint,showpacs,showkeys,amsmath,amssymb]{revtex4}
\usepackage{graphicx} 
\usepackage{color}
\usepackage{ulem} 
\newcommand{\prycatri}{(Pr$_{1-y}$Y$_{y}$)$_{0.7}$Ca$_{0.3}$CoO$_3$}
\newcommand{\prcaco}{Pr$_{1-x}$Ca$_{x}$CoO$_3$}

\newcommand{\lasrco}{La$_{1-x}$Sr$_{x}$CoO$_3$}
\newcommand{\lacaco}{La$_{1-x}$Ca$_{x}$CoO$_3$}
\newcommand{\prcapet}{Pr$_{0.5}$Ca$_{0.5}$CoO$_3$}

\newcommand{\lasrpet}{La$_{0.5}$Sr$_{0.5}$CoO$_3$}

\newcommand{\prcatri}{Pr$_{0.7}$Ca$_{0.3}$CoO$_3$}
\newcommand{\ndcatri}{Nd$_{0.7}$Ca$_{0.3}$CoO$_3$}

\newcommand{\laco}{LaCoO$_3$}
\newcommand{\lnco}{$Ln$CoO$_3$}

\newcommand{\ndco}{NdCoO$_3$}
\newcommand{\st}{$\rm ^o$}
\newcommand{\etal}{\textit{et al.}}
\definecolor{bkat}{rgb}{1.0,0.0,0.0}
\definecolor{bkk}{rgb}{0.0,0.0,1.0}
\begin{document}
\sloppy
\title{Metal-insulator transition and the Pr$^{3+}$/Pr$^{4+}$ valence shift in
  (Pr$_{1-y}$Y$_{y}$)$_{0.7}$Ca$_{0.3}$CoO$_3$}
\author{J. Hejtm\'{a}nek}
\author{E. \v{S}antav\'{a}}
\author{K. Kn\'{\i}\v{z}ek}
\author{M. Mary\v{s}ko}
\author{Z. Jir\'{a}k}
\affiliation{
 Institute of Physics ASCR, Cukrovarnick\'a 10, 162 00 Prague 6, Czech Republic. \\
}
\author{T. Naito}
\author{H. Sasaki}
\author{H. Fujishiro}
\affiliation{
 Faculty of Engineering, Iwate University, 4-3-5 Ueda, Morioka 020-8551, Japan. \\
}
\begin{abstract}
The magnetic, electric and thermal properties of the ($Ln_{1-y}$Y$_{y}$)$_{0.7}$Ca$_{0.3}$CoO$_3$
perovskites  ($Ln$~=~Pr, Nd) were investigated down to very low temperatures. The main attention
was given to a peculiar metal-insulator transition, which is observed in the praseodymium based
samples with $y=0.075$ and 0.15 at $T_{M-I}=64$ and 132~K, respectively. The study suggests that
the transition, reported originally in Pr$_{0.5}$Ca$_{0.5}$CoO$_3$, is not due to a mere change of
cobalt ions from the intermediate- to the low-spin states, but is associated also with a
significant electron transfer between Pr$^{3+}$ and Co$^{3+}$/Co$^{4+}$ sites, so that the
praseodymium ions occur below $T_{M-I}$ in a mixed Pr$^{3+}$/Pr$^{4+}$ valence. The presence of
Pr$^{4+}$ ions in the insulating phase of the yttrium doped samples
(Pr$_{1-y}$Y$_{y}$)$_{0.7}$Ca$_{0.3}$CoO$_3$ is evidenced by Schottky peak originating in Zeeman
splitting of the ground state Kramers doublet. The peak is absent in pure
Pr$_{0.7}$Ca$_{0.3}$CoO$_3$ in which metallic phase, based solely on non-Kramers Pr$^{3+}$ ions,
is retained down to the lowest temperature.

\end{abstract}
\maketitle

\section{Introduction}

Thermally induced transitions in \lnco\ ($Ln$~=~La, Y, rare earths) have been studied for decades.
Recent interpretation associates them with a local excitation of the octahedrally coordinated
Co$^{3+}$ ions from LS (low spin, $t_{2g}^6$) to HS (high spin, $t_{2g}^4 e_g^2$) state, followed
at higher temperature by a formation of a metallic phase of IS character (intermediate spin,
$t_{2g}^{5} \sigma^*$) - see \textit{e.g.} \cite{RefJirak2008PRB78_014432} and citations therein.
An attention is attracted also to the mixed-valence cobaltites like \lasrco\ or \lacaco\, where a
similar transition from insulating LS Co$^{3+}$ ground state towards the metallic $t_{2g}^{5}
\sigma^*$ one is obtained in the course of doping ($0<x<0.2$). As concerns the metallic region
beyond $x=0.2$, some heavily doped Pr-based cobaltites, in particular \prcapet\, behave
anomalously. At ambient temperature, they appear in the metallic $t_{2g}^{5} \sigma^*$ phase as
expected, but on cooling they undergo a sharp M-I (metal-insulator) transition at T$_{M-I}=90$~K,
documented for the first time by S.~Tsubouchi \etal\
\cite{RefTsubouchi2002PRB66_052418,RefTsubouchi2004PRB69_144406}. The same transition was observed
also on the less-doped samples \prcaco\ ($x=0.3$) under high pressures or upon a partial
substitution of praseodymium by smaller rare earth cations or yttrium
\cite{RefFujita2004JPSJ73_1987,RefFujita2005JPSJ74_2294,RefNaito2010JPSJ79_034710}. The effect was
tentatively ascribed to a change of the cobalt states from itinerant ones $t_{2g}^{5} \sigma^*$ to
a mixture of localized LS Co$^{3+}$ ($t_{2g}^6$, S~=~0) and LS Co$^{4+}$ ($t_{2g}^5$, S~=~1/2)
states. Very recently, an alternative explanation has been proposed based on electronic structure
calculations and some indirect experimental data, namely the significant lattice contraction and
shortening of Pr-O bond lengths that accompany the M-I transition
\cite{RefKnizek2010PRB81_155113,RefBaronGonzalez2010PRB81_054427}. It is suggested that the formal
cobalt valency in \prcapet\ is changed at T$_{M-I}$ from the mixed-valence Co$^{3.5+}$ towards
pure Co$^{3+}$ with strong preference for LS state, and the praseodymium valence is simultaneously
increased from Pr$^{3+}$ towards Pr$^{4+}$. The spin state transition and formation of an
insulating state in \prcapet\ is thus an analogy of the compositional transition from
ferromagnetic metal \lasrpet\ to diamagnetic insulator \laco\
\cite{RefWu2003PRB67_174408,RefKnizek2010JMMM322_1221}.

The present work concerns the system \prycatri\ ($y=0-0.15$), which is advantageous with respect
to an easier preparation of oxygen stoichiometric samples as compared with the \prcapet\ system.
The M-I transition observed for $y\geq0.075$ ($T_{M-I}=60-130$~K) is manifested by pronounced
anomalies in the temperature course of transport, magnetic and thermal properties. Fundamental
information on the nature of the insulating phase is obtained at very low temperatures. This
refers especially to the occurrence of a Schottky peak in the specific heat, centered at $1-2$~K.
A comparison with the specific heat data in \ndcatri\, where similar Schottky peak is currently
also observed, allows us to conclude that the peak arises due to an exchange field splitting of
the doublet ground state of Kramers rare-earth ions, in particular Nd$^{3+}$. Its occurrence in
the yttrium doped \prycatri\ provides a direct and quantitative evidence for the fact that in the
insulating phase, in addition to common Pr$^{3+}$ (non-Kramers ion),  there is a significant
significant population of Pr$^{4+}$ (Kramers ion).

\section{Experimental}

Polycrystalline samples of ($Ln_{1-y}$Y$_{y}$)$_{0.7}$Ca$_{0.3}$CoO$_3$ ($Ln$~=~Pr, Nd) were
prepared by a solid-state reaction. Raw powders of Pr$_6$O$_{11}$, Nd$_2$O$_3$, Y$_2$O$_3$,
Co$_3$O$_4$, and CaCO$_3$ were weighted with proper molar ratios and ground using an agate mortar
and pestle for 1~h. Mixed powders were calcined at 1000\st C for 24~h in air. Then they were
pulverized, ground and pressed into pellets of 20 mm diameter and 4 mm thickness. Pellets were
sintered at 1200\st C for 24~h in 0.1~MPa flowing oxygen gas. The measured densities of each
sample were greater than 90~\% of the ideal density. Powder X-ray diffraction patterns were taken
for each sample using CuK$_\alpha$ radiation; the samples were confirmed to have a single phase
orthoperovskite ($Pbnm$) structure.

The magnetic measurements  were performed in the temperature range from 2  to 400 K using a SQUID
magnetometer (MPMS-XL). The hysteresis loops at T~=~2 and 4.5 K were recorded between the field -7
and 7~T. The susceptibility was measured under an applied field of 0.1 T, employing the zero
field- (ZFC) and field-cooled  (FCC) regimes during warming and cooling the sample,  respectively.

Thermal conductivity, thermoelectric power and electrical resistivity were measured using a
four-probe method with a parallelepiped sample cut from the sintered pellet. The electrical
current density varied depending on the sample resistivity between $10^{-1}$~A/cm$^2$ (metallic
state) and $10^{-7}$~A/cm$^2$ (insulating state). The measurements were done on sample cooling and
warming using a close-cycle cryostat working down to $2-3$~K. The detailed description of the cell
including calibration is described elsewhere \cite{RefHejtmanek1999PRB60_14057}.

The specific heat was measured by PPMS device (Quantum Design) using the two$-\tau$ model. The
data were collected generally on sample cooling. The experiments at very low temperatures (down to
0.4~K) were done using the He$^3$ option.

\section{Results}

\subsection{Physical characterization}

\begin{figure}[h]
\includegraphics[width=0.90\columnwidth,viewport=5 420 580 780,clip]{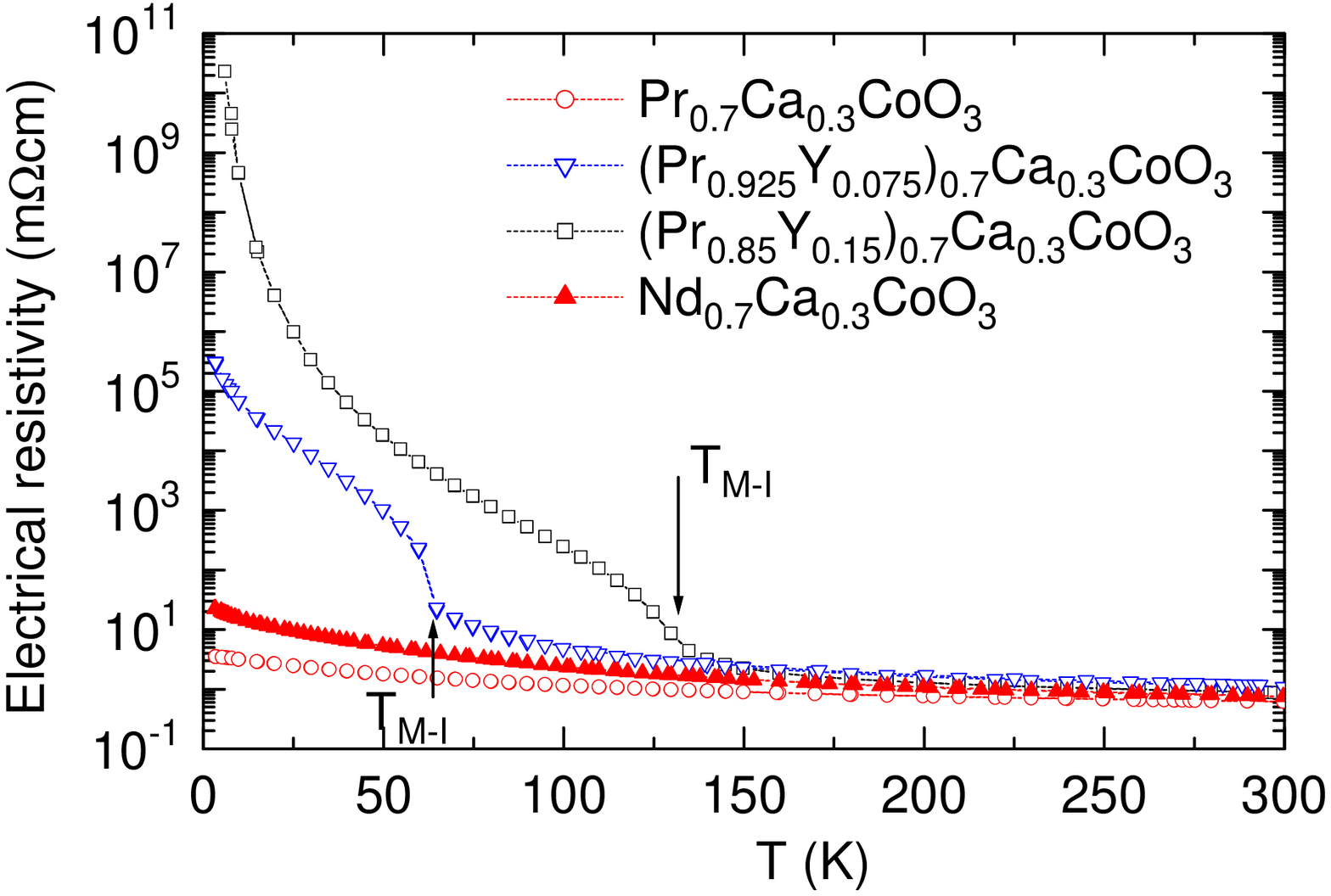}
\includegraphics[width=0.90\columnwidth,viewport=5 420 580 780,clip]{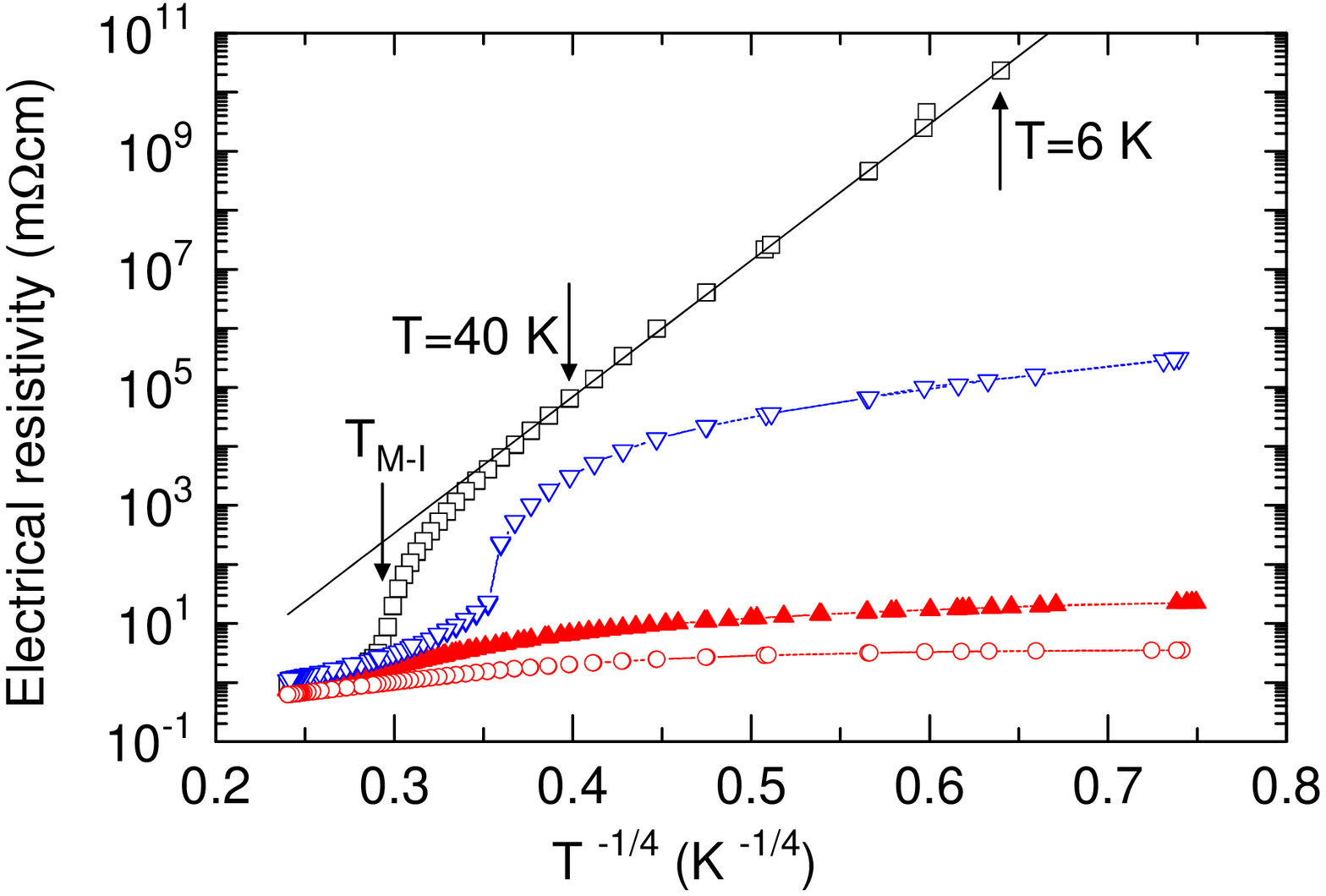}
\caption{(Color online) The temperature dependence of electrical resistivity in \prycatri\ ($y=0$,
0.075 and 0.15) and \ndcatri. The data measured on cooling and warming are overlapping. The lower
panel shows the dependence on $T^{-1/4}$.} \label{FigER1}
\end{figure}

\begin{figure}[h]
\includegraphics[width=0.90\columnwidth,viewport=5 380 580 780,clip]{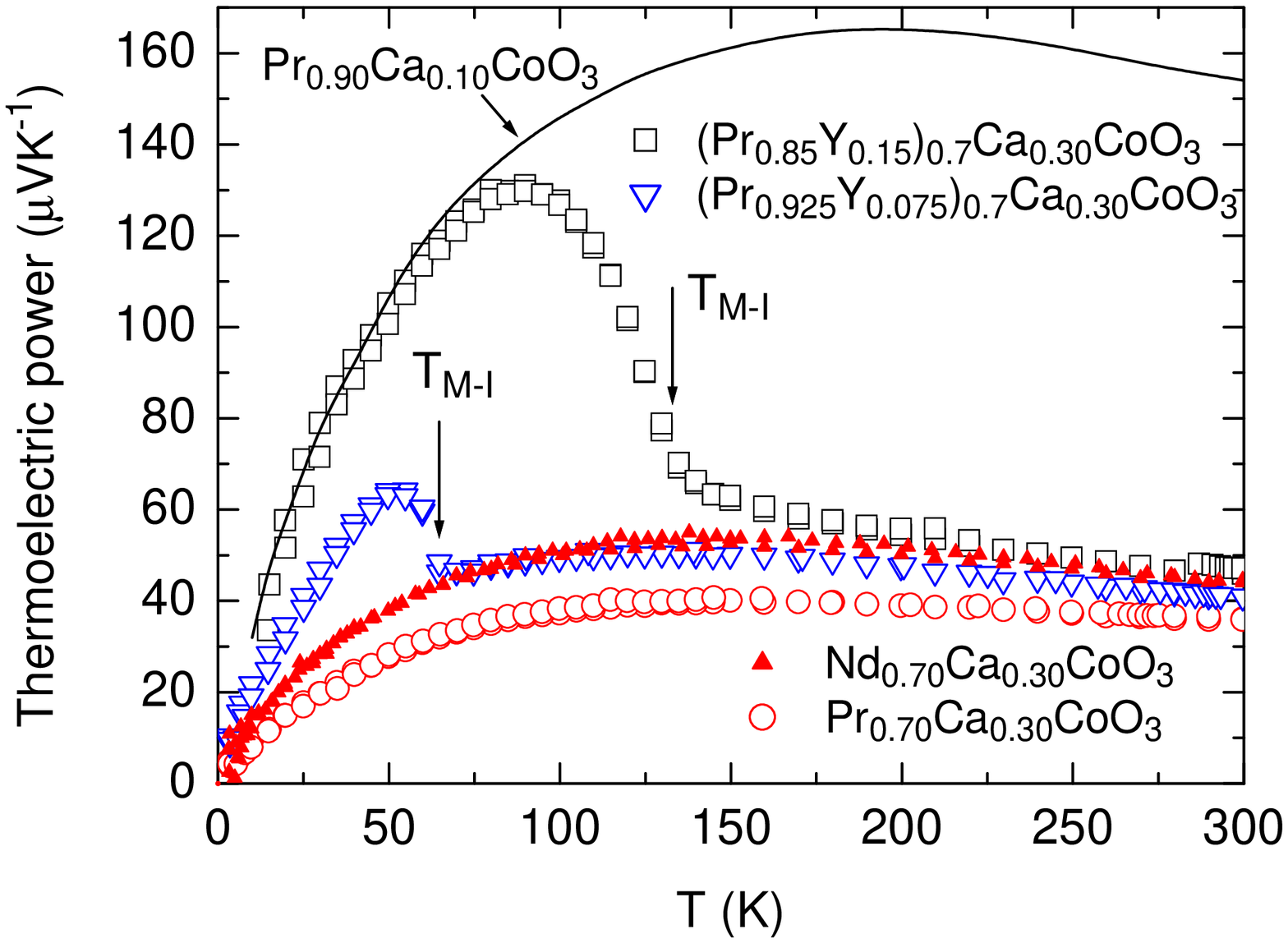}
\caption{(Color online) The enhanced Seebeck coefficient in \prycatri\ ($y=0.075$ and 0.15) below
the M-I transition. The data for metallic \prcatri\ and \ndcatri\ are also shown. The solid line
gives data for Pr$_{0.9}$Ca$_{0.1}$CoO$_3$ taken of Ref. \cite{RefMasuda2003JPSJ72_873} (see the
text). } \label{FigS}
\end{figure}

\begin{figure}[h]
\includegraphics[width=0.90\columnwidth,viewport=5 380 580 780,clip]{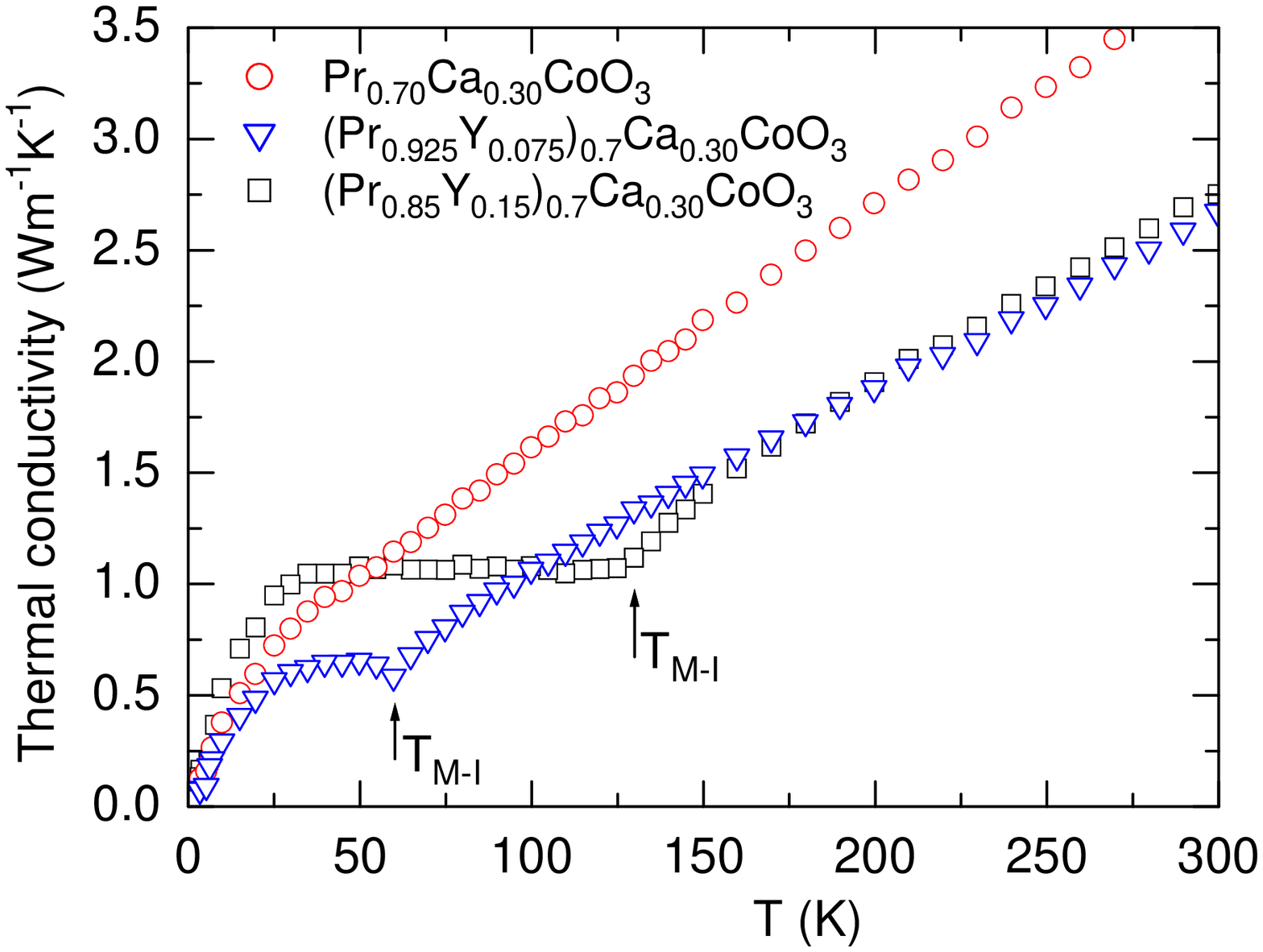}
\caption{(Color online) Thermal conductivity of \prycatri\ for $y=0$, 0.075 and 0.15.}
\label{FigTC}
\end{figure}

\begin{figure}[h]
\includegraphics[width=0.90\columnwidth,viewport=5 420 580 780,clip]{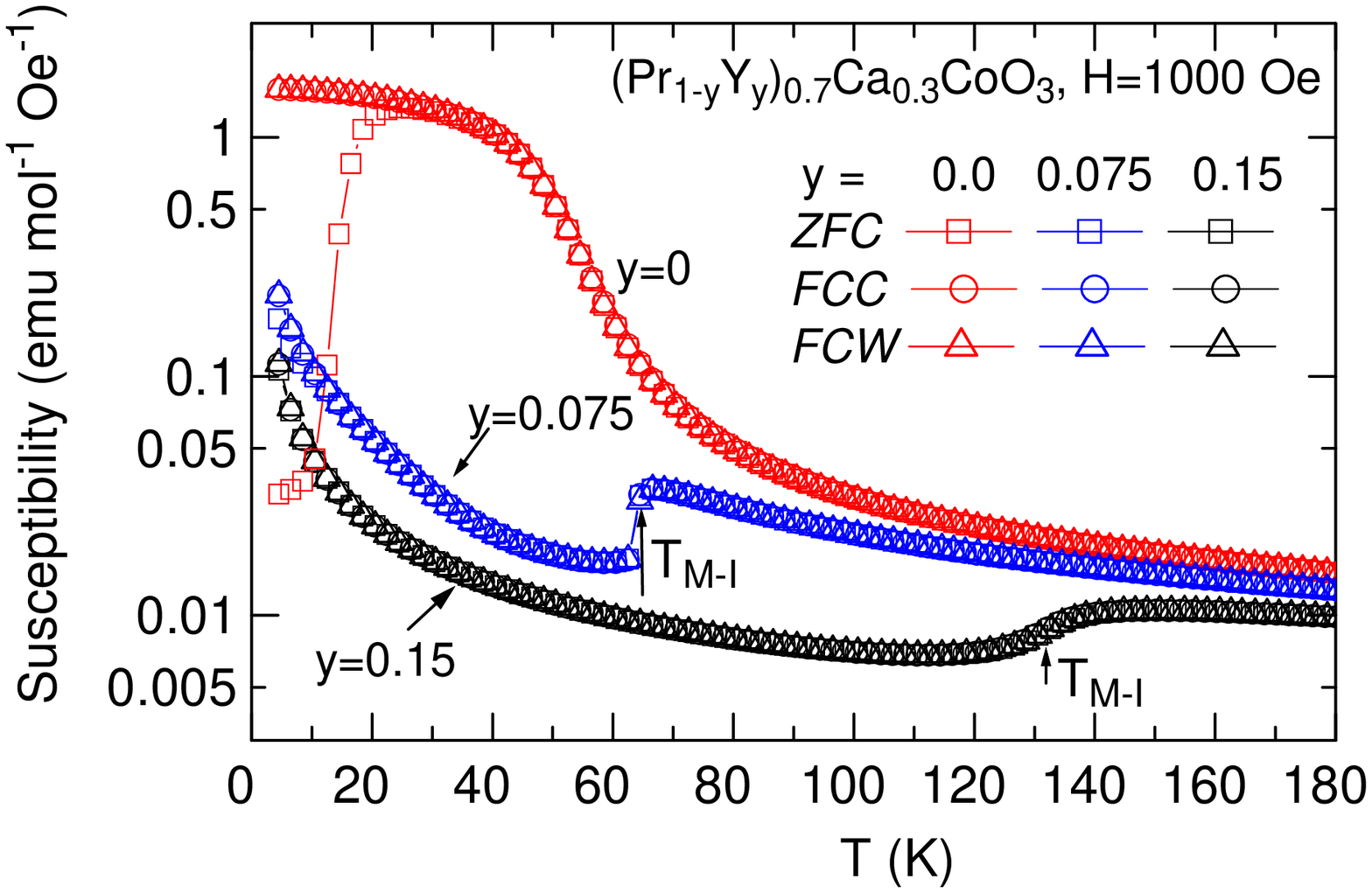}
\includegraphics[width=0.90\columnwidth,viewport=5 420 580 780,clip]{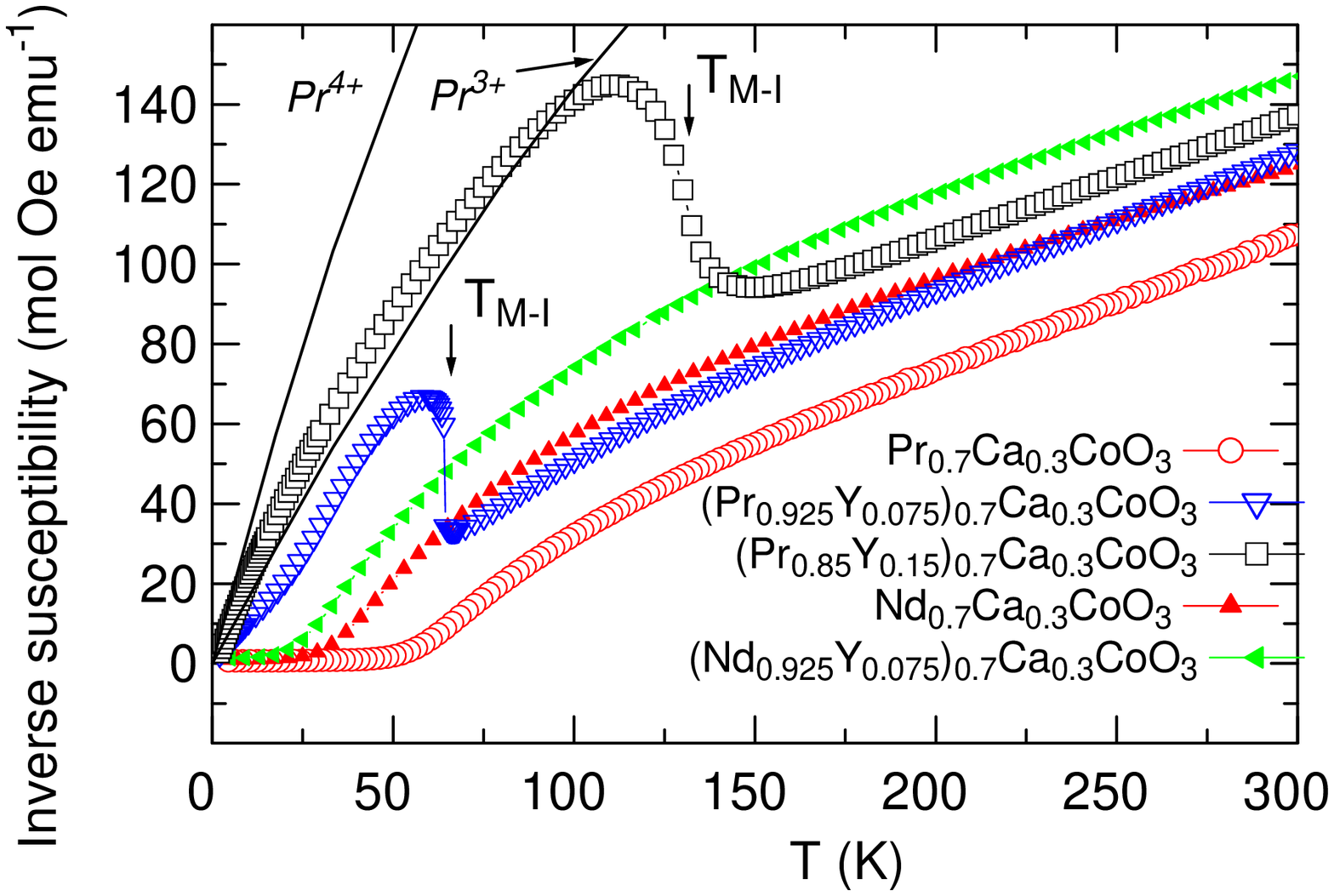}
\caption{(Color online) Magnetic susceptibility of \prycatri\ for $y=0$, 0.075 and 0.15 (in $log$
scale). The lower panel shows the temperature dependence of inverse susceptibility. The solid
lines mark the theoretical contribution for Pr$^{3+}$ and Pr$^{4+}$ ions, calculated for the
praseodymium amount in $y=0.15$ using the data in Ref. \cite{RefSekizawa1998JMMM177_541} (see the
text). The data for (Nd$_{1-y}$Y$_{y}$)$_{0.7}$Ca$_{0.3}$CoO$_3$ for $y=0$ and 0.075 without M-I
transition are added for comparison.} \label{FigMS}
\end{figure}

\begin{figure}[h]
\includegraphics[width=0.90\columnwidth,viewport=5 380 580 780,clip]{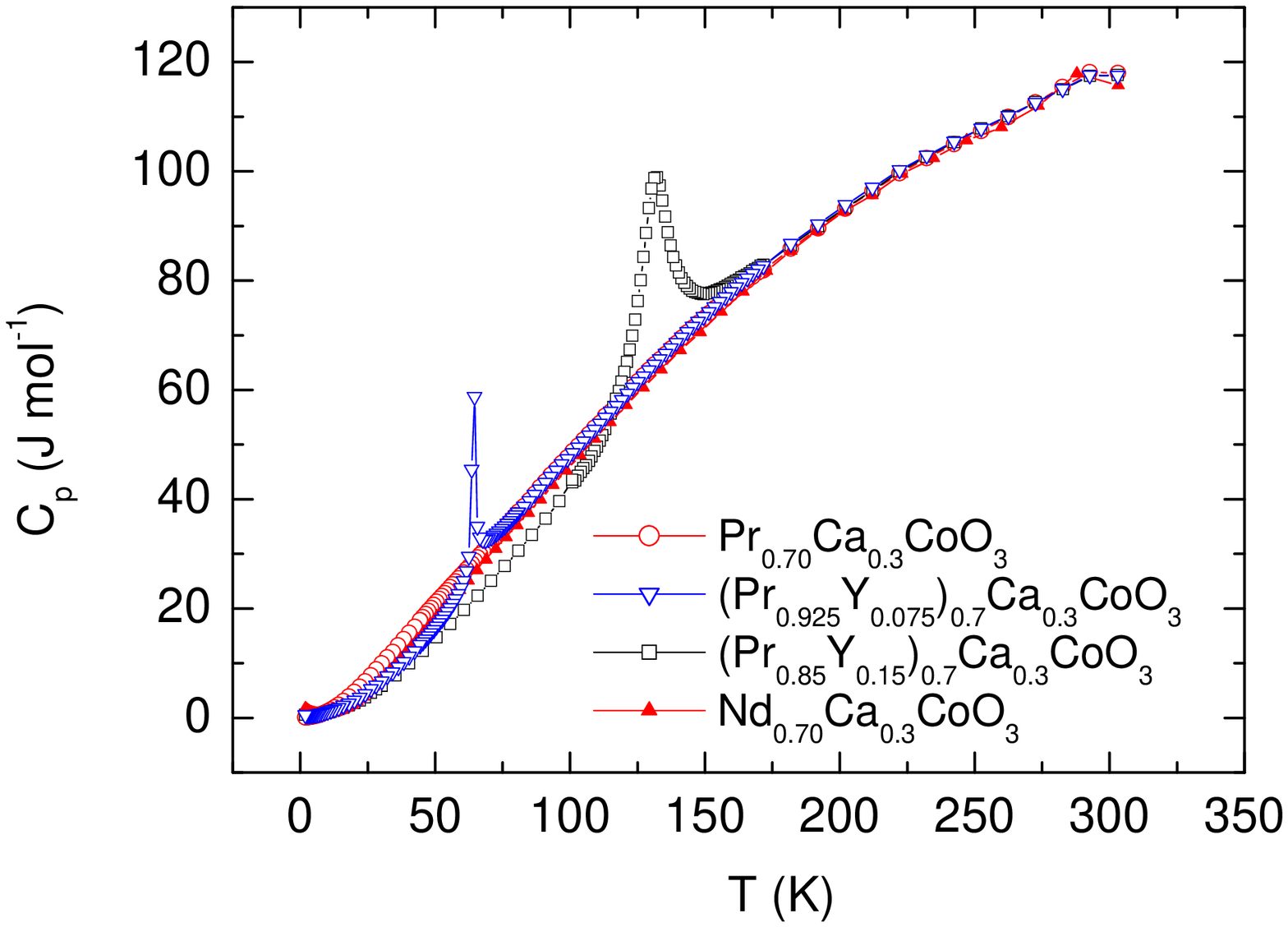}
\caption{(Color online) Specific heat of \prycatri\ for $y=0$, 0.075 and 0.15. The data for
\ndcatri\ are added for comparison. } \label{FigHC}
\end{figure}

The electric transport measurements on the \prycatri\ samples for $y=0$, 0.075 and 0.15 and on
\ndcatri\ are presented in Figs.~\ref{FigER1} and \ref{FigS}. It is seen that the $y=0$ and
analogous neodymium sample are metallic over the whole temperature range, tending to a finite
resistivity of about 1~m$\Omega$cm at zero K. The yttrium doped samples, apparently in the same
metallic state at room temperature, exhibit a sudden increase of the electrical resistivity and
thermopower (Seebeck coefficient) at $T_{M-I}=64$~K and 132~K for $y=0.075$ and $y=0.15$,
respectively. Concurrently, the thermal conductivity exhibits a kink, followed with a recovery at
lower temperatures (Fig.~\ref{FigTC}). The magnetic susceptibility drops markedly
(Fig.~\ref{FigMS}), which is a strong signature that cobalt ions transform to LS states at
$T_{M-I}$. The M-I transition in \prycatri\ is further manifested by a pronounced peak in the
specific heat data (Fig.~\ref{FigHC}). The peak is very narrow $\sim1-2$~K for $y=0.075$ with
lower $T_{M-I}$, suggesting a first order character of the transition, while it is much broader
$\sim15$~K for $y=0.15$ with higher $T_{M-I}$. It is seen also that the values of specific heat
below the M-I transition are small compared to \prcatri\ retaining the metallic phase, which is
indicative of a significant change of lattice dynamics in the low-temperature phase of yttrium
doped samples.

\begin{figure}[h]
\includegraphics[width=0.90\columnwidth,viewport=5 380 580 780,clip]{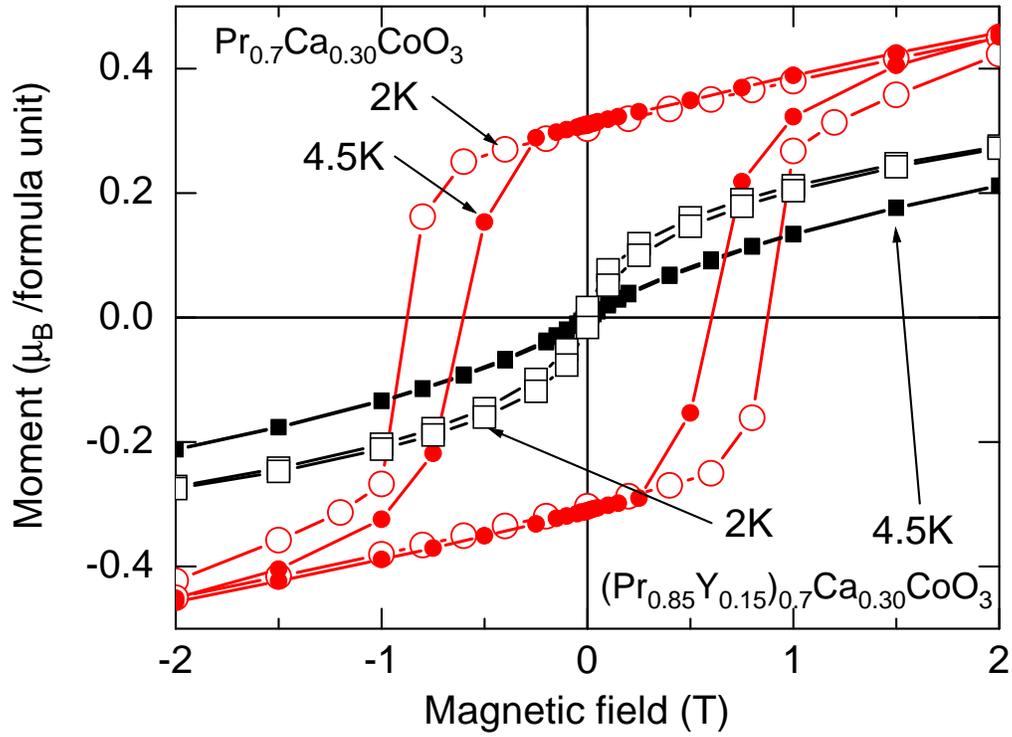}
\caption{(Color online) Magnetization loops measured at 2~K (open symbols) and 4.5~K (solid
symbols) on \prycatri\ ($y=0$ and 0.15).} \label{FigHL}
\end{figure}

The anomalies in the transport, magnetic and thermal data, similar to those in
Figs.~\ref{FigER1}-\ref{FigHC}, were observed earlier for the prototypical compound \prcapet\
\cite{RefTsubouchi2002PRB66_052418,RefTsubouchi2004PRB69_144406} and also for the \prcatri-derived
systems with analogous M-I transition. In important distinction to these previous reports, the
transition in present samples \prycatri\ exhibits very small thermal hysteresis, which is about
0.2~K for both $y=0.075$ and 0.15. Another property deserving attention is practical absence of
residual metallic phase in the insulating ground state, especially for $y=0.15$. As seen in the
plot of inverse susceptibility in lower panel of Fig.~\ref{FigMS}, the metallic sample \prcatri\
undergoes FM-like transition at $T_C\sim55$~K and, consistently, the magnetization data taken at
low temperatures show nearly rectangular hysteresis loops with large coercivity
(Fig.~\ref{FigHL}). On the other hand, the sample $y=0.15$ exhibits essentially paramagnetic
behavior of inverse susceptibility and magnetization curves are of Brillouin type, though some
very weak coercivity and remanence appear at 2~K.

The very low susceptibility below $T_{M-I}$ makes sample $y=0.15$ suitable for a more quantitative
analysis. The observed values are obviously smaller than the calculated contribution of
praseodymium ions in trivalent state (see data of Ref. \cite{RefSekizawa1998JMMM177_541} in lower
panel of Fig.~\ref{FigMS}), disregarding that there should also be contribution of the LS
Co$^{4+}$ spins. A plausible interpretation provides the new model of M-I transition stating that
some praseodymium ions are changed to tetravalent states with lower magnetic moments, and
corresponding number of LS Co$^{4+}$ ions are transformed to diamagnetic LS Co$^{3+}$. The actual
valence shift can be inferred from the thermopower data in Fig.~\ref{FigS}. Namely, the
thermopower in mixed-valence cobaltites is primarily determined by number of carriers,
\textit{i.e.} by the formal cobalt valence, and, in a less extent, by the average ionic size of
large cations in perovskite A-sites. Having this in mind, we refer to the fact that Seebeck
coefficient in the metallic phase of yttrium doped sample $y=0.15$ ($T\sim150~-~300$~K) matches
well to \ndcatri\ with comparable A-site size. This means that cobalt valence in these two samples
is practically the same and, consequently, praseodymium ions are essentially in trivalent state,
similarly to neodymium ones. (We estimate Pr$^{3.03+}$ as upper possible limit.) Below
$T_{M-I}=132$~K, Seebeck coefficient is enhanced and, on further cooling, it becomes comparable to
samples with much lower number of carriers, in particular to Pr$_{0.9}$Ca$_{0.1}$CoO$_3$
\cite{RefMasuda2003JPSJ72_873}. In a rough estimate, this may mean that the formal cobalt valence
in $y=0.15$ is changed upon the M-I transition from original Co$^{3.3+}$ to about Co$^{3.1+}$.

\subsection{The low-temperature specific heat}

\begin{figure}[h]
\includegraphics[width=0.90\columnwidth,viewport=5 330 580 780,clip]{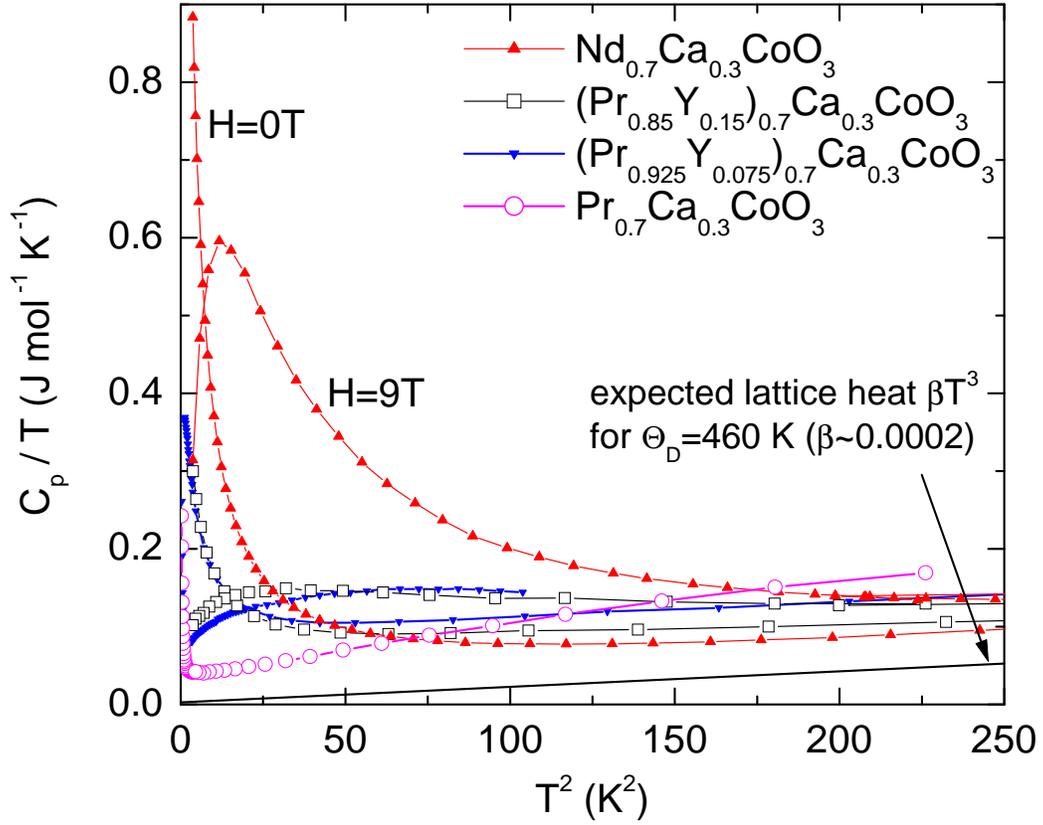}
\caption{(Color online) The low-temperature specific heat of \prycatri\ and \ndcatri\ shown in the
c$_p/T$ $vs.$ $T^2$ plot. Beside common cubic lattice term $\beta T^3$ and linear $\gamma T$ term,
one may notice an additional contribution for $y=0$, associated with excitations within the
crystal field split multiplet of Pr$^{3+}$, and an important Schottky anomaly for $y=0.075, 0.15$
and \ndcatri\ that shifts with external field.} \label{FigHC2}
\end{figure}

At the lowest temperatures, the insulating $y=0.075$ and 0.15 samples exhibit a steep increase of
specific heat (see the c$_p/T$ $vs.$ $T^2$ plot in Fig.~\ref{FigHC2}). Though a similar effect was
reported originally for \prcapet\, the present experiments, performed down to 0.4~K, document for
the first time that the increase is due to Schottky peak, centered at $T=1-2$~K, that adds to
common lattice, electronic and nuclear contributions of low-temperature specific heat
(Fig.~\ref{FigLT}). With increasing external field the peak position shifts rapidly to higher
temperatures. The Schottky peak is absent in metallic \prcatri\, but is found with even larger
intensity in \ndcatri\ with analogous metallic ground state (see Figs.~\ref{FigHC2} and
\ref{FigndVLT}).

The occurrence of Schottky peak in \ndcatri\ can be understood considering the Kramers character
of the rare earth ions. In the orthoperovskite structure, the $^4$I$_{9/2}$ multiplet of Nd$^{3+}$
is split by crystal field effects into five relatively distant doublets
\cite{RefPodlesnyak1993JPCM5_8973}. The double degeneracy is lifted by an exchange field arising
due to FM ordering of cobalt spins. Hence, the Schottky peak under discussion is related to
thermal excitations within the ground state doublet, the location of the maximum defines the
energy splitting due to exchange field ($\Delta = 0.42~k_BT_{max}$) and its shift with applied
field bears information on the effective gyromagnetic $g_{J'}$ value that governs the Zeeman
splitting ($\Delta = g_{J'} J'\mu_B(B+B_{m})$, where J'=$\pm1/2$ denotes two pseudospin levels of
the Kramers doublet and $B_{m}$ is molecular field acting on the rare-earth moment). The shape of
observed Schottky peak is nearly ideal with only a little extra broadening, and the integration of
$c_{Schottky}/T$ over $T$ gives the total change of entropy of 3.95~JK$^{-1}$mol$^{-1}$, in good
agreement with the theoretical value 0.7N$k_B ln2$=4.04~JK$^{-1}$mol$^{-1}$ for \ndcatri\
composition.

In the case of \prcatri\, the $^3$H$_4$ multiplet of Pr$^{3+}$ is split into nine singlets
\cite{RefPodlesnyak1994JPCM6_4099}. In accordance with the non-magnetic (singlet) ground state,
there is no Schottky peak at very low temperatures but another Schottky-like contribution emerges
at $T>10$~K, as can be seen by increased specific heat of \prcatri\ in Figs.~\ref{FigHC2} and
\ref{FigLT}. It originates in an excitation to the next singlet state at an energy difference of
5~meV.

Analogously to \ndcatri\, the Schottky peak in the yttrium doped samples \prycatri\ can be related
to the presence of Kramers ions, which are Pr$^{4+}$ states formed presumably below $T_{M-I}$. A
splitting of the $^2$F$_{5/2}$ multiplet to three doublets with large spacing is anticipated. The
total entropy change is determined from data in Fig.~\ref{FigLT} to 0.98~JK$^{-1}$mol$^{-1}$ for
$y=0.15$. This quantitative analysis enables to estimate the number of Pr$^{4+}$ states to about
0.18 per formula. For $y=0.075$, the entropy change makes 0.61~JK$^{-1}$mol$^{-1}$ and
corresponding number of Pr$^{4+}$ states is 0.12 per formula.

The change of Zeeman splitting of the ground state doublets with external field is presented in
Fig.~\ref{FigShift}. There is a little shift of $T_{max}(Schottky)$ for \ndcatri\ between zero
field and 3~T, before it starts to rise nearly linearly. This shows that the external field does
not simply add to the molecular field, in other words, there must be antiparallel or perpendicular
orientation of the Nd$^{3+}$ moments with respect to the FM polarized Co spins in the ground
state. On the other hand, the shift with external field in the $y=0.15$ sample \prycatri\ is
monotonous from the very beginning, which is indicative of a parallel orientation of the Pr$^{4+}$
and Co moments.

\section{Discussion}

As established in earlier works, the M-I transition in heavily doped cobaltites, like \prcapet\ or
\prycatri\, is conditioned by presence of praseodymium ions, combined with a suitable structural
distortion which depends on an average ionic radius and size mismatch of the perovskite A-site
ions (see \cite{RefNaito2010JPSJ79_034710} and references therein). Important finding of the
recent $GGA+U$ electronic structure calculations for Pr$_{0.5}$Ca$_{0.5}$CoO$_3$ is the location
of the occupied Pr$^{3+}$-~4f states closely below E$_F$ at ambient temperature, and their
splitting and partial transfer above E$_F$ due to lattice contraction experimentally observed at
$T_{M-I} =90$~K. In ionic picture, this means that praseodymium valence in the metallic phase is
Pr$^{3+}$, and Pr$^{4+}$ valence states are formed in the low-temperature insulating phase,
compensated by a valence shift of cobalt ions towards pure Co$^{3+}$. Such scenario is currently
supported using different experiments on \prycatri\, namely the low-temperature specific heat,
thermopower and magnetic susceptibility. In addition, some new information on physical properties
over a broad temperature range is obtained and deserves more discussion.

The first issue is the character of electronic transport. The data in Fig.~\ref{FigER1} suggest
that the resistivity in pure \prcatri\ tends to a finite value. The metallicity is thus evident
despite of the granular character of the ceramic sample. On the other hand, the $y=0.15$ sample of
\prycatri\ shows a clear localization that is best fitted by Mott's formula for variable range
hopping (VRH), $\rho=\rho_o.exp(To/T)^{1/4}$, valid for $T<40K$ (see Fig.~\ref{FigER1}b). The
characteristic parameters are $\rho_o\sim4\times10^{-5}$ m$\Omega$cm and $T_o\sim8\times10^6$~K.
The VRH mechanism is associated with a phonon-assisted tunneling of electrons from initial sites
located near E$_F$ to target sites close in energy, for which the Miller-Abrahams transfer rate
applies (see \textit{e.g.} recent review of N.~Tessler \etal\
\cite{RefTessler2009AdvMater21_2741}). This type of conduction is generally manifested with a
specific $T^{-1/2}$ dependence of thermopower, which is, however, not obeyed for present samples.
Instead, the Seebeck coefficient increases steeply from zero value at the lowest temperatures in a
linear metallic-like manner and then tends to a saturation, which is not reached completely
because of ingoing transition. Such behavior is suggesting that firstly, the present system
possesses a quasi-continuous, very narrow band of electronic levels at E$_F$, and secondly, the
Seebeck coefficient is related to the presence of carriers rather than to their motion,
\textit{i.e.} the dominant contribution is the change of the net entropy of a solid upon the
addition of a charge carrier, while the energy transported by carriers, divided by the absolute
temperature, seems to be marginal - see \textit{e.g.} Ref.~\cite{RefEminThermoelectricsHandbook}.

Major experimental data refer, however, to specific heat in the $y=0.075$ and 0.15 samples with
occurrence of the M-I transition, compared to samples \prcatri\ and \ndcatri\ with metallic ground
state (Figs.~\ref{FigHC} and~\ref{FigHC2}). The dominant contribution to the specific heat above
the liquid helium temperature is the phononic term that follows in the metallic samples a standard
Debye-type dependence, though there is also extra contribution due to thermally activated
population of the crystal field split levels of the rare earths (see \textit{e.g.}
\cite{RefKnizek2009PRB79_134103}). At room temperature the c$_p$ value reaches
118~JK$^{-1}$mol$^{-1}$ for all the studied samples. This makes about 93\% of the saturated
lattice heat taking into account the Dulong-Petit high temperature limit
c$_V$=15N$k_B\sim$125~JK$^{-1}$mol$^{-1}$ in ABO$_3$ perovskites and a minor correction for
thermal expansion c$_p$-c$_V\sim$1.2~JK$^{-1}$mol$^{-1}$ at 300~K
\cite{RefKnizek2009PRB79_134103}.

The phononic term in the low-temperature phase of the yttrium doped \prycatri\ samples is clearly
lower than that of the undoped samples. This is a strong indication that M-I transition is
accompanied by a significant change of orthoperovskite structure, presumably by the lattice
contraction and Pr-O bond length shortening, similar to those in \prcapet.

As concerns the specific heat anomaly at T$_{M-I}$, the associated entropy change is determined,
by integration of an excess of c$_p/T$, to 2.17 and 4.78~JK$^{-1}$mol$^{-1}$ for \prycatri\ with
$y=0.075$ and 0.15, respectively. It is of interest that the total entropy change at the gradual
spin-state transitions in \lnco\ is about 20 JK$^{-1}$mol$^{-1}$ and is distributed evenly between
the first step from the LS~Co$^{3+}$ groundstate to a LS/HS mixture and the subsequent formation
of the metallic phase of IS~Co$^{3+}$ character
\cite{RefKnizek2009PRB79_134103,RefTachibana2008PRB77_094402}.

\begin{figure}
\includegraphics[width=0.90\columnwidth,viewport=5 380 580 780,clip]{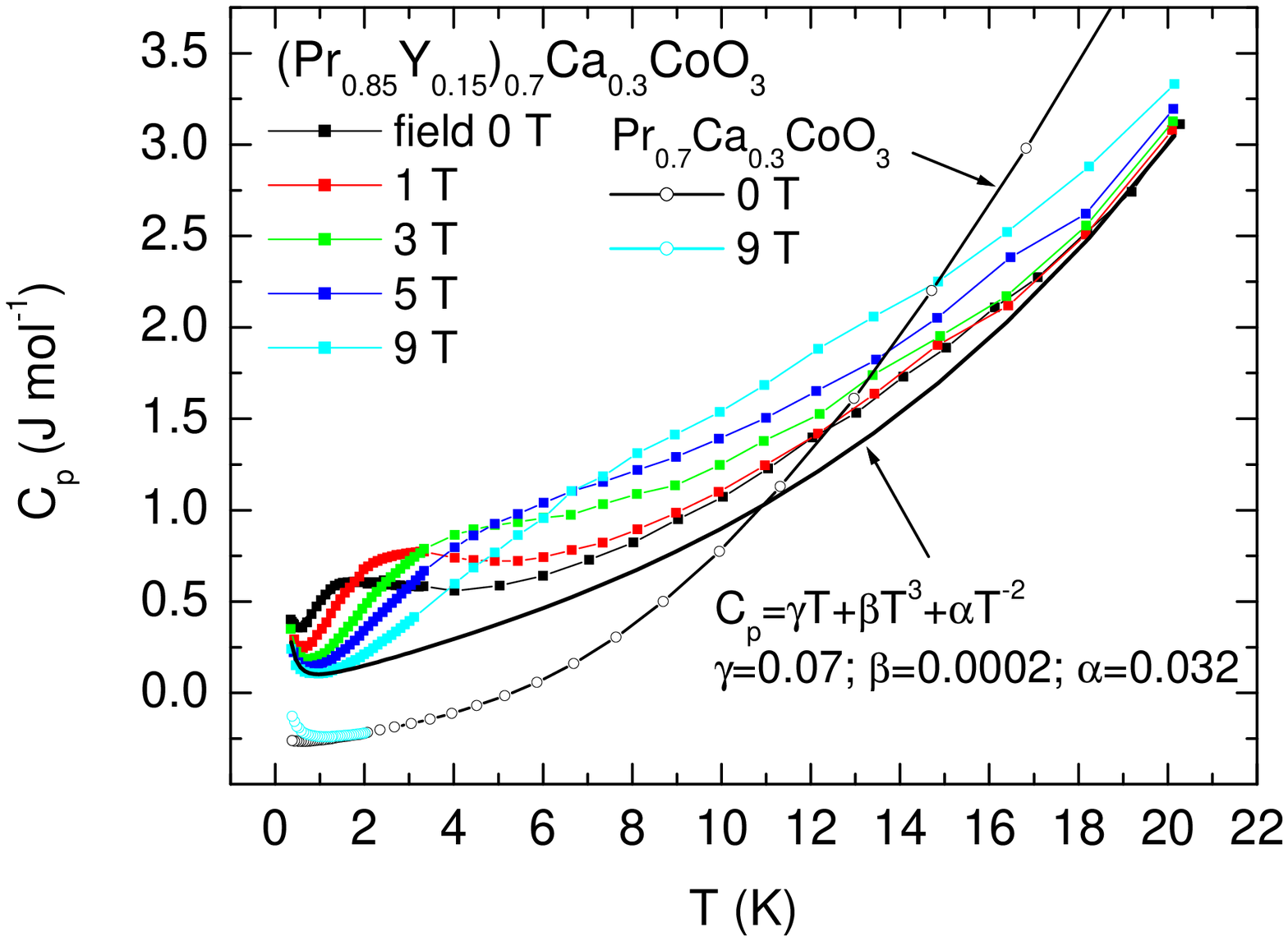}
\caption{(Color online) Specific heat of \prycatri\ ($y=0.15$) down to 0.4~K, measured in fields
$0-9$~Tesla. The heat capacity for metallic \prcatri\ ($y=0$) added for comparison is displaced to
-0.3.} \label{FigLT}
\end{figure}

\begin{figure}
\includegraphics[width=0.90\columnwidth,viewport=5 380 580 780,clip]{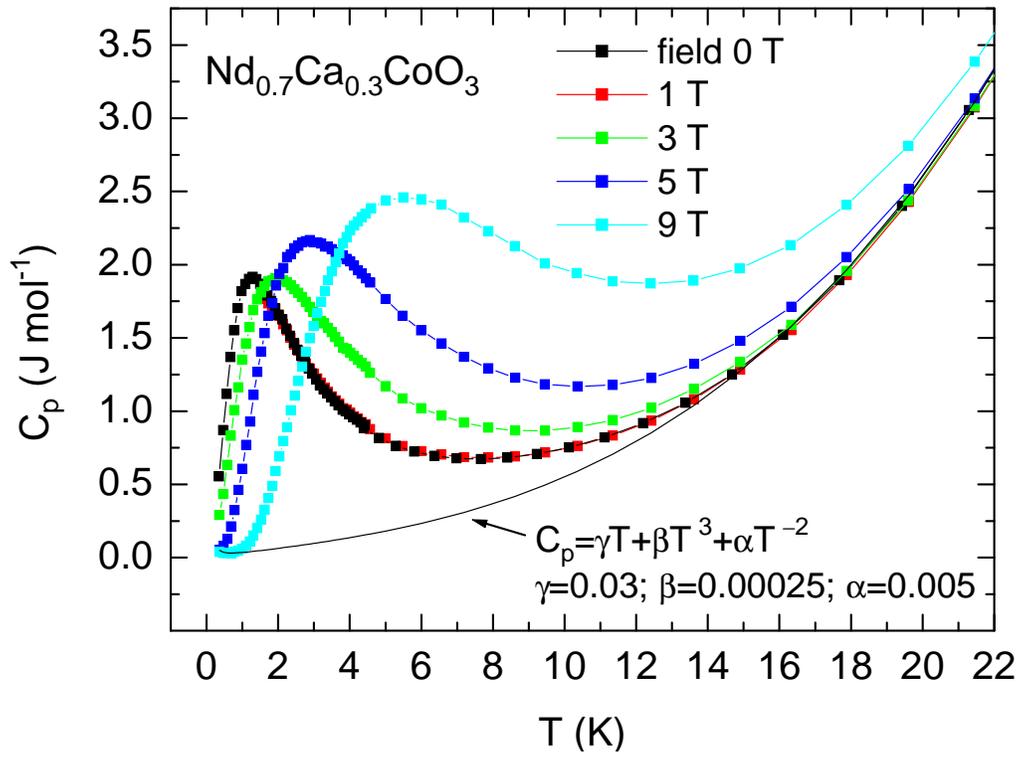}
\caption{(Color online) Specific heat of \ndcatri\ down to 0.4~K, measured in fields 0~-~9 Tesla.}
\label{FigndVLT}
\end{figure}

At low-enough temperatures, there are two standard contributions of the specific heat, that can be
easily distinguished in the c$_p/T$ $vs.$ $T^2$ plot in Fig.~\ref{FigHC2}. One is the vanishing
phononic term, approximated by the cubic term $\beta T^3$ where $\beta$ is directly related to
Debye temperature $\Theta_D$. The parameters $\beta\sim0.00020$ and 0.00024~JK$^{-4}$mol$^{-1}$
for the $y=0.15$ sample and \ndcatri\, shown in the fit in Figs.~\ref{FigLT} and \ref{FigndVLT},
give an estimate $\Theta_D=460$ and 430~K for the insulating and metallic phase, respectively. The
latter value should be compared with $\Theta_D\sim 400$~K determined for metallic
La$_{0.7}$Sr$_{0.3}$CoO$_3$ \cite{RefHe2009PRB80_214411}. The second contribution is the linear
term $\gamma T$. The $\gamma$ value $\sim 30~$mJK$^{-2}$mol$^{-1}$ for metallic \prcatri\ and
\ndcatri\ is comparable to $\sim 40~$mJK$^{-2}$mol$^{-1}$ for La$_{0.7}$Sr$_{0.3}$CoO$_3$
\cite{RefHe2009PRB80_214411}. The $\gamma$ parameter is, however, still larger in the insulating
phase of the yttrium doped samples \prycatri, which corroborate the idea of charge carriers
squeezed in a very narrow band. As concerns the baseline in Fig.~\ref{FigLT} and \ref{FigndVLT},
the remaining term is the nuclear contribution, manifested as the $\alpha T^{-2}$ upturn at the
lowest temperature. The value $\alpha\sim 0.032~$mJK$^{-3}$mol$^{-1}$, observed on the $y=0.15$
sample \prycatri\ is exceptionally large and apparently field independent. It is worth mentioning
that recent specific heat study on \lasrco\ report much smaller nuclear specific heat despite the
bulk ferromagnetic state \cite{RefHe2009PRB80_214411}. In these compounds, the nuclear heat
originates in the Zeeman splitting of the spin $I=7/2$ multiplet of $^{59}$Co nuclei in the
hyperfine field induced by FM ordering of electronic spins. Its intensity thus probes the amount
of FM phase in the sample. The very large $\alpha T^{-2}$ term in the $y=0.15$ sample should be
thus ascribed primarily to contribution of $^{141}$Pr nuclei with spin $I=5/2$ in the hyperfine
field induced by dressed Pr$^{4+}$ electronic pseudospins. On the other hand, the specific heat
data for pure \prcatri\, also included in Fig.~\ref{FigLT}, do not show observable nuclear
contribution at zero field, but a comparable $\alpha T^{-2}$ term is induced in field of 9~T.

\begin{figure}
\includegraphics[width=0.90\columnwidth,viewport=5 380 580 780,clip]{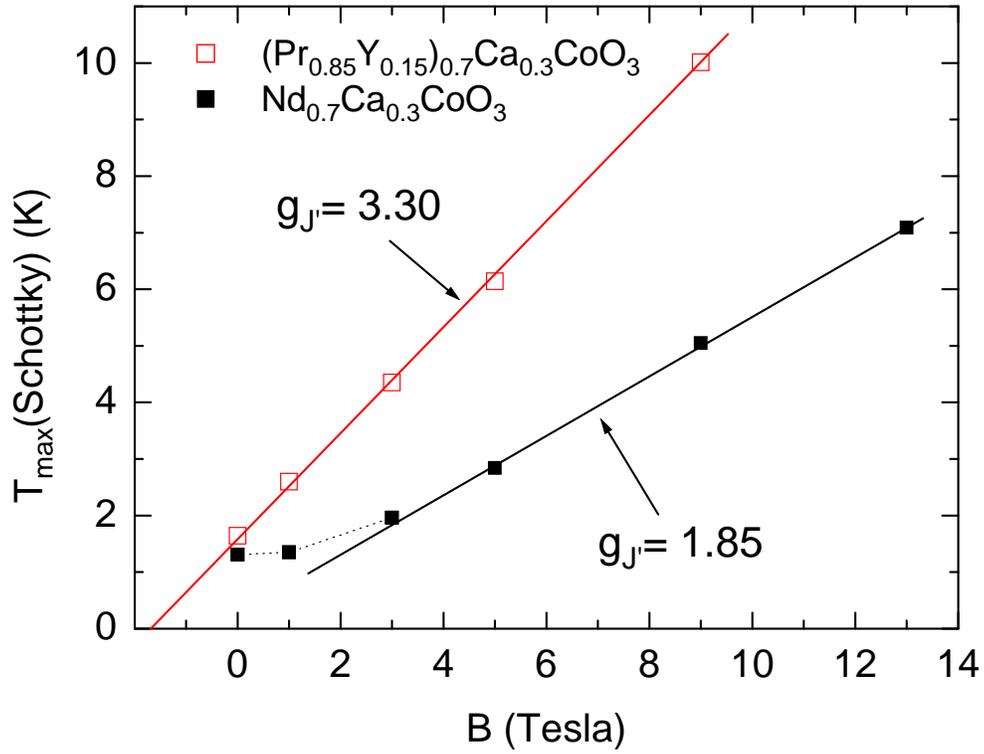}
\caption{(Color online) The shift of the Schottky peak maximum ($T_{max}$ in the
c$_{Schottky}~vs.~T$ dependence) with applied field. The lines define the effective g-factors of
the Kramers ground state doublets of Pr$^{4+}$ and Nd$^{3+}$. The molecular field experienced by
rare-earth moments at $B=0$ is determined to $\sim1.6$~Tesla in \prycatri\ ($y=0.15$) and
$\sim2.5$~Tesla in \ndcatri.} \label{FigShift}
\end{figure}

The Schottky peaks observed in \ndcatri\ (Fig.~\ref{FigLT}) and yttrium doped \prycatri\
(Fig.~\ref{FigndVLT}) serve as local probe of the Kramers ions Nd$^{3+}$ and Pr$^{4+}$. Their
position in the zero-field specific heat and shift with applied field allow to estimate that
Nd$^{3+}$ pseudospins in \ndcatri\ are characterized by effective $g_{J'}=1.85$ and experience a
molecular field of $B_{m}=2.5$~T, while for Pr$^{4+}$ pseudospins in the $y=0.15$ sample
\prycatri\ the respective values are $g_{J'}=3.30$ and $B_{m}=1.6$~T (see Fig.~\ref{FigShift}).
The origin of molecular field in metallic \ndcatri\ is in ferromagnetic ordering of cobalt ions in
IS states at $T_C\sim25$~K ($M_s\sim0.16~\mu_B$ per f.u.) and their interaction via $3d-4f$
exchange mechanisms with spin component of the Nd$^{3+}$ moments. The situation in the
low-temperature phase of \prycatri\ which exhibits essentially paramagnetic characteristics (see
Figs.~\ref{FigMS} and~\ref{FigHL}) is not clear and will require further experimental and
theoretical investigation. If Pr$^{4+}$  pseudospins were ordered spontaneously, significantly
sharper and higher peak (so called lambda peak) would be observed as is \textit{e.g.} in the case
of antiferromagnetic ordering of Nd$^{3+}$ moments in \ndco\ at $T_N = 1.2$~K
\cite{RefBartolome1994SSC91_177}. The observation of standard Schottky peak thus suggests that
also in the yttrium doped samples the rare earths experience a stable molecular field formed in
the cobalt subsystem. We note that the cobalt subsystem in the $y=0.15$ sample contains 12\% of
LS~Co$^{4+}$ ions that represent spins 1/2 in a diamagnetic matrix of LS~Co$^{3+}$. Anticipating
their itinerancy they may lead to certain ferromagnetic polarization of the narrow $t_{2g}$ cobalt
bands, low compared to \ndcatri\ but sufficient to mediate relatively strong magnetic interactions
among the Pr$^{4+}$ pseudospins (RKKY model known for $3d-4f$ intermetallics can be envisaged
\cite{RefFranseRadwanski}). Such interpretation seems to be supported by two findings - the
parallel orientation of praseodymium moments with cobalt spins, manifested in the field-induced
shift of Schottky peak in Fig.~\ref{FigShift}, and the unusually large term $\alpha T^{-2}$
contributed by the $^{141}$Pr nuclear spins, seen for the $y=0.15$ sample \prycatri\ in
Fig.~\ref{FigLT}.

\section{Conclusion}

A comparative study of perovskite cobaltites ($Ln_{1-y}$Y$_{y}$)$_{0.7}$Ca$_{0.3}$CoO$_3$
perovskites ($Ln$~=~Pr, Nd) was undertaken with an aim to elucidate the character of a peculiar
first-order M-I transition in some Pr-based cobaltites. Though the transition is typical for
\prcapet, the present yttrium doped systems \prycatri\ with $T_{M-I}=64$ and 132~K for $y=0.075$
and 0.15, respectively, appear preferable because of easier stabilization of the stoichiometric
phase and complete transformation from the metallic to insulating state. The study shows that the
M-I transition is manifested by a huge peak in the specific heat data and marked changes in the
electrical resistivity, thermopower and thermal conductivity. A sudden drop of magnetic
susceptibility indicates a change of cobalt states from the metallic $t_{2g}^{5} \sigma^*$ to the
mixture of LS~Co$^{3+}$ ($t_{2g}^6$) and LS~Co$^{4+}$ ($t_{2g}^5$).

An important novelty is an observation of Schottky peak in specific heat data at very low
temperatures, $0.4-10$~K. This peak is absent in pure \prcatri\ in which metallic phase is
retained down to the lowest temperature, but appears with large intensity in \ndcatri\ with
similar metallic phase. Its occurrence follows from the Kramers character of Nd$^{3+}$ and
Pr$^{4+}$ whose $^4$I$_{9/2}$ and $^2$F$_{5/2}$ multiplets are split by crystal field associated
with the distorted dodecahedral coordination of the rare earths in the $Pbnm$ perovskites. The
Schottky peak thus probes the Zeeman splitting of the ground state doublet. The total entropy
change associated with the Schottky peak is $k_B ln2$ per ion, which allows to determine the
concentration of Kramers ions in the samples experimentally, by integration of $c_{Schottky}/T$
over $T$. The analysis for the $y=0.075$ and 0.15 samples \prycatri\ provides values 0.12 and 0.18
Pr$^{4+}$ per formula unit, respectively. Considering that praseodymium ions are essentially in
trivalent state in the high-temperature phase, the observation of Pr$^{4+}$ in the low-temperature
phase is in accordance with idea that the simultaneous M-I and spin-state transition in Pr-based
cobaltites is accompanied by an electronic transfer between the praseodymium and cobalt ions. In
particular for $y=0.15$, the present results are indicative of a significant change from common
valence distribution in the metallic state,
(Pr$^{3+}_{0.85}$Y$^{3+}_{0.15}$)$_{0.7}$Ca$^{2+}_{0.3}$Co$^{3+}_{0.7}$Co$^{4+}_{0.3}$O$^{2-}_{3}$,
to
(Pr$^{3+}_{0.59}$Pr$^{4+}_{0.26}$Y$^{3+}_{0.15}$)$_{0.7}$Ca$^{2+}_{0.3}$Co$^{3+}_{0.88}$Co$^{4+}_{0.12}$O$^{2-}_{3}$
in the insulating state.

\textbf{Acknowledgments}. This work was supported by Project No.~202/09/0421 of the Grant Agency
of the Czech Republic.



\end{document}